\documentclass[prb,aps,twocolumn,showpacs,ams,superscriptaddress]{revtex4-1}
\usepackage{graphicx}
\usepackage{dcolumn}
\usepackage{latexsym}
\usepackage{color}
\usepackage{bm}

\begin{document}

% Use the \preprint command to place your local institutional report
% number in the upper righthand corner of the title page in preprint mode.
% Multiple \preprint commands are allowed.
% Use the 'preprintnumbers' class option to override journal defaults
% to display numbers if necessary
%\preprint{}

%Title of paper
\title{Spin-polarized local density of states \\
in the vortex state of helical $p$-wave superconductors}

% repeat the \author .. \affiliation  etc. as needed
% \email, \thanks, \homepage, \altaffiliation all apply to the current
% author. Explanatory text should go in the []'s, actual e-mail
% address or url should go in the {}'s for \email and \homepage.
% Please use the appropriate macro foreach each type of information

% \affiliation command applies to all authors since the last
% \affiliation command. The \affiliation command should follow the
% other information
% \affiliation can be followed by \email, \homepage, \thanks as well.

\author{Kenta K. Tanaka} 
\email[]{ktanaka@mp.okayama-u.ac.jp}
\affiliation{
Department of Physics, Okayama University, Okayama 700-8530, JAPAN}

\author{Masanori Ichioka}
\email[]{ichioka@cc.okayama-u.ac.jp}
\affiliation{
Department of Physics, Okayama University, Okayama 700-8530, JAPAN}
\affiliation{
Research Institute for Interdisciplinary Science, Okayama University, Okayama 700-8530, JAPAN}

\author{Seiichiro Onari} 
\affiliation{
Department of Physics, Okayama University, Okayama 700-8530, JAPAN}
\affiliation{
Research Institute for Interdisciplinary Science, Okayama University, Okayama 700-8530, JAPAN}

%Collaboration name if desired (requires use of superscriptaddress
%option in \documentclass). \noaffiliation is required (may also be
%used with the \author command).
%\collaboration can be followed by \email, \homepage, \thanks as well.
%\collaboration{}
%\noaffiliation

\date{\today}

\begin{abstract}
Properties of the vortex state in helical $p$-wave superconductor are studied by the quasi-classical Eilenberger theory.
We confirm the instability of the helical $p$-wave state at high fields and that 
the spin-polarized local density of states $M(E, {\bm r})$ appears even when Knight shift does not change.
This is because the vorticity couples to the chirality of up-spin pair or down-spin pair of the helical state.
In order to identify the helical $p$-wave state at low fields,
we investigate the structure of the zero-energy $M(E=0, {\bm r})$ in the vortex states,
and also the energy spectra of $M(E, {\bm r})$.
\end{abstract}

% insert suggested PACS numbers in braces on next line
%\pacs{pacs ???}
%\pacs{74.25.Uv, 74.20.Rp, 74.25.nj, 74.25.Ha}
%74.	Superconductivity (for superconducting devices, see 85.25.-j)
%74.20.-z	Theories and models of superconducting state
%74.20.Pq	Electronic structure calculations (for methods of electronic structure calculations, see 71.15.-m)
%74.20.Rp	Pairing symmetries (other than s-wave)
%74.25.Ha	Magnetic properties including vortex structures and related phenomena (for vortices, magnetic bubbles, and magnetic domain structure, see 75.70.Kw)
%74.25.Jb	Electronic structure (photoemission, etc.)
%74.25.nj	Nuclear magnetic resonance
%74.25.Op	Mixed states, critical fields, and surface sheaths
%74.25.Uv	Vortex phases (includes vortex lattices, vortex liquids, and vortex glasses)
%74.25.Wx	Vortex pinning (includes mechanisms and flux creep)
%74.62.En	Effects of disorder%
%74.70.Pq	Ruthenates

% insert suggested keywords - APS authors don't need to do this
%\keywords{}

%\maketitle must follow title, authors, abstract, \pacs, and \keywords
\maketitle

%%%%%%%%%%%%%%%%%%%%%%%%%%%%%%%%%%%%%%%%%%%%%%%%%%%%%
%-------------------------------------------  Introduction   -------------------------------------------  
%%%%%%%%%%%%%%%%%%%%%%%%%%%%%%%%%%%%%%%%%%%%%%%%%%%%%
%\section{INTRODUCTION}
%
\section{Introduction}
The superconductor (SC) ${\rm{Sr_2RuO_4}}$ has attracted much attention as a topological SC,
since exotic quantum states such as a Majorana state are expected in the vortex and surface states.
A lot of experimental and theoretical studies support that ${\rm{Sr_2RuO_4}}$ is a spin-triplet chiral $p$-wave SC
~\cite{Sr2RuO4-1, Sr2RuO4-2}.
On the other hand,
the helical $p$-wave state also has been suggested
as another scenario~\cite{Rice-Sigrist, Takamatsu_GL, Scaffidi}.
This is because the detailed structure of $d$-vector in ${\rm{Sr_2RuO_4}}$ remains unclear.
In addition, the difference of condensation energy between chiral and helical states is very small
compared to the transition temperature~\cite{Tsuchiizu}.
Therefore, 
we need methods to distinguish between chiral and helical states in experiments 
for  ${\rm{Sr_2RuO_4}}$ or other candidate materials for spin-triplet SC.
For the purpose,
it is necessary that 
we study a unique behavior of physical quantity depending on the symmetry of $d$-vector.

In the bulk state of chiral SC,
the time-reversal symmetry is broken
because of the angular momentum of Cooper pair $L_z{\neq}0$.
The chirality of chiral $p$-wave state, i.e., $L_z={\pm}1$ can be distinguished 
via coherence effect in the vortex state.
In fact,
previous theories suggested that the impurity effects on the local density of states (LDOS) and 
local NMR relaxation rate $T_1^{-1}$ show different behaviors between $p_+$ and $p_-$ states~\cite{Kato-Hayashi, Tanuma, Kurosawa, Hayashi-NMR, K.Tanaka3}.
This chirality dependence is caused by the interaction between the chirality and the vorticity,
depending on whether the chirality is parallel ($L_z=+1$) or anti-parallel ($L_z=-1$) to the vorticity $(W=+1)$~\cite{Ichioka-chiral, Heeb}.
On the other hand, 
in the bulk state of helical $p$-wave SC,
the time-reversal-invariant superconductivity appears
since $L_z={\pm}1$ are quenched
with the degeneracy between up-spin and down-spin pairs.
The up-spin (down-spin) pair's order-parameter 
${\Delta}_{{\uparrow}{\uparrow}} ({\Delta}_{{\downarrow}{\downarrow}})$ characterized by $S_z=+1 (-1)$ 
has chirality $L_z=-1 (+1)$ so that the bulk condition $L_z+S_z=0$~\cite{Rice-Sigrist}.
Therefore,
in the vortex state of helical $p$-wave SC,
spin states of low-energy excitations may show a unique behavior, 
reflecting the vorticity coupling to the chirality of 
${\Delta}_{{\uparrow}{\uparrow}}(L_z=-1)$ or ${\Delta}_{{\downarrow}{\downarrow}}(L_z=+1)$.

The scanning tunneling microscopy and spectroscopy (STM/STS) measurement can directly detect
the LDOS via excitations in the vortex state~\cite{Hess, Fischer}.
Recently,
the STM/STS measurement
in the vortex state of topological insulator-superconductor ${\rm{Bi_2Te_3/NbSe_2}}$ heterostructure
has performed~\cite{STM_Majonara1},
and theoretical studies for the measurement
have supported the existence of Majorana zero-energy mode in the vortex core
~\cite{Hetero-theory, STM_Majonara2}.
Moreover, spin polarization of Majorana zero-energy modes are investigated 
by the spin-polarized STM/STS measurement,
which can selectively detect the spin-dependent conductance~\cite{SPSTM}.
The spin polarization in the vortex state of topological SC ${\rm{Cu}}_x{\rm{Bi_2Si_3}}$ is also theoretically studied~\cite{Nagai}.

In this paper,  
we study properties of the helical $p$-wave SC,
and focus on the spin-polarized LDOS in the vortex lattice state,
in order to reveal a unique behavior of the helical state.
In particular,
we calculate the structure of the zero-energy spin-polarized LDOS at low fields,
and also the energy spectra.
These results help to investigate the vortex state of helical $p$-wave SC and Majorana zero-energy state 
by spin-polarized STM/STS measurement.

This paper is organized as follows.
After the introduction, we describe our formulation of the quasi-classical Eilenberger equation in the vortex lattice state and 
the calculation method for the spin-resolved LDOS in Sec. II.
In Sec. III, we investigate the $H$-dependence of order-parameter,
and examine the instability of the helical state at high fields.
In Sec. IV, we show the $H$-dependence of the zero-energy spin-polarized DOS and LDOS.
The $E$-dependence of the spin-polarized LDOS is presented in Sec. V.
The last section is devoted to the summary.

%%%%%%%%%%%%%%%%%%%%%%%%%%%%%%%%%%%%%%%%%%%%%%%%%%%%%
%-------------------------------------------  Formulation   -------------------------------------------  
%%%%%%%%%%%%%%%%%%%%%%%%%%%%%%%%%%%%%%%%%%%%%%%%%%%%%
%\section{FORMULATION}
%
\section{Formulation}
We calculate the spatial structure of vortices in the vortex lattice state by quasi-classical Eilenberger theory.
The quasi-classical theory is valid when the atomic scale is small enough compared to the superconducting coherence length.
For many SCs including ${\rm Sr_2RuO_4}$,
the quasi-classical condition is well satisfied~\cite{Sr2RuO4-1, Sr2RuO4-2}.
Moreover,
since our calculations are performed in the vortex lattice state,
we can obtain the structure of LDOS quantitatively. 

For simplicity, we consider the helical $p$-wave pairing 
on the two-dimensional cylindrical Fermi surface, 
${\bm k}=(k_x,k_y)=k_{\rm F}(\cos\theta_k,\sin\theta_k) $, 
and the Fermi velocity 
${\bm v}_{\rm F}=v_{\rm F0} {\bm k}/k_{\rm F}$. 
In the following,
the symbol of hat indicates the $2{\times}2$ matrix in spin space and
the symbol of check indicates the $4{\times}4$ matrix in particle-hole and spin spaces.

To obtain quasi-classical Green's functions
%---Elenberger eq.---------
${\check{g}}({\rm i}\omega_n, {\bm r},{\bm k})$
in the vortex lattice state, 
we solve Riccati equation derived from Eilenberger equation~\cite{Tsutsumi}
\begin{eqnarray} 
-{\rm i} {\bm{v}} {\cdot} {\nabla}{\check{g}}({\rm i}\omega_n,{\bm r},{\bm k}) =
{\frac{1}{2}} [ {\rm i} {\tilde{\omega}}_n  {\check{\sigma}}_z - {\check{\Delta}}({\bm r},{\bm k})  ,  {\check{g}}({\rm i}\omega_n,{\bm r},{\bm k})] \ \ \
\label{eq:Eil}
\end{eqnarray} 
in the clean limit, where  
${\bm r}$ is the center-of-mass coordinate of the pair, 
${\bm v}={\bm v}_{\rm F}/v_{{\rm F}0}$,
${\check{\sigma}}_z$ is the Pauli matrix,
and
${\rm i} {\tilde{\omega}}_n = {\rm i} {\omega}_n - {\bm{v}} {\cdot} {\bm{A}}$ 
with Matsubara frequency ${\omega}_n$.
The quasi-classical Green's function and order parameter are described by 
\begin{eqnarray}
&{\check{g}}( {\rm i} {\omega}_n,{\bm r},{\bm k})= -{\rm {i}} {\pi}
\left[
\begin{array}{cc}
{\hat{g}}( {\rm i} {\omega}_n,{\bm r},{\bm k}) & {\rm i} {\hat{f}}( {\rm i} {\omega}_n,{\bm r},{\bm k}) \\
-{\rm i} {\underline{\hat{f}}}( {\rm i} {\omega}_n,{\bm r},{\bm k}) & -{\hat{g}}( {\rm i} {\omega}_n,{\bm r},{\bm k})
\end{array}
\right], \ \ \ \\
&{\check{\Delta}}({\bm r},{\bm k})= 
\left[
\begin{array}{cc}
0 & {\hat{\Delta}} ({\bm r},{\bm{k}}) \\
-{\hat{\Delta}}^{\dagger} ({\bm r},{\bm{k}}) & 0
\end{array}
\right]
\end{eqnarray}
where ${\check{g}}^2 = - {\pi}^2 {\check{1}}$.
The spin spaces of ${\hat{g}}$ and ${\hat{\Delta}}$ are defined by the matrix elements 
$g_{{\sigma}{\sigma}'} ( {\rm i} {\omega}_n,{\bm r},{\bm k})=  
[g_0 ( {\rm i} {\omega}_n,{\bm r},{\bm k}) {\hat{1}} + {\sum_{{\mu}=x,y,z}} g_{\mu}( {\rm i} {\omega}_n,{\bm r},{\bm k}) {\hat{\sigma}}_{\mu}]_{{\sigma}{\sigma}'}$ and
${\Delta}_{{\sigma}{\sigma}'} ({\bm r},{\bm{k}}) = [{\rm i} {\sum_{{\mu}=x,y,z}}  (   d_{\mu} ({\bm r},{\bm k}) {\cdot} {\hat{\sigma}}_{\mu}    ) {\hat{\sigma}}_y]_{{\sigma}{\sigma}'}$
where ${\sigma}, {\sigma}' =$ ${\uparrow}$(up-spin) or ${\downarrow}$(down-spin), 
and $d_{\mu}$ is $\mu$-component of $d$-vector.
In addition, 
the matrix elements of order-parameter are defined by
\begin{eqnarray}
{\Delta}_{{\sigma}{\sigma}'} ({\bm r},{\bm{k}})= 
{\Delta}_{+,{\sigma}{\sigma}'}({\bm r}) \phi_{p+}({\bm k}) +   {\Delta}_{-,{\sigma}{\sigma}'}({\bm r}) \phi_{p-}({\bm k})
\label{eq:op-def} 
\end{eqnarray}
with the order-parameter ${\Delta}_{{\pm},{\sigma}{\sigma}'}({\bm r})$
and pairing function ${\phi}_{p_{\pm}}({\bm k}) =k_x {\pm} {\mathrm{i}}k_y$ for $p_{\pm}$-state.
Length, temperature, and magnetic field are, 
respectively, measured in unit of 
$\xi_0$, $T_{\rm c}$, and $B_0$. 
Here, $\xi_0=\hbar v_{\rm F0}/2 \pi k_{\rm B} T_{\rm c}$, 
$B_0=\phi_0 /2 \pi \xi_0^2$ with the flux quantum $\phi_0$.  
$T_{\rm c}$ is superconducting transition temperature at 
a zero magnetic field.  
The energy $E$, pair potential $\Delta$ and $\omega_n$ 
are in unit of $\pi k_{\rm B} T_{\rm c}$. 
In the following, we set $\hbar=k_{\rm B}=1$. 
In this study, our calculations are performed at $T=0.5T_c$. 

We set the magnetic field along the $z$ axis.
The vector potential   
${\bm A}({\bm r})=\frac{1}{2} {\bm H} \times {\bm r}
 + {\bm a}({\bm r})$ in the symmetric gauge.
${\bm H}=(0,0,H)$ is a uniform flux density, 
and ${\bm a}({\bm r})$ is related to the internal field 
${\bm B}({\bm r})=(0,0,B({\bm r})) 
={\bm H}+\nabla\times {\bm a}({\bm r})$.
The unit cell of the vortex lattice is set as square lattice~\cite{Sr2RuO4-1}.

%%%%%%%%%%%%%%%%%%%%%%%%%%%%%%%%%%%%%%%%%%%%%%%%%%%%%
%-------------------------------------------  Self-consistent   -------------------------------------------  
%%%%%%%%%%%%%%%%%%%%%%%%%%%%%%%%%%%%%%%%%%%%%%%%%%%%%

To determine the pair potential ${\hat{\Delta}} ({\bm r})$ and 
the quasi-classical Green's functions selfconsistently, 
we calculate the order-parameter ${\hat{\Delta}}_{\pm} ({\bm r})$ by the gap equation
\begin{eqnarray}
{\hat{\Delta}}_{\pm} ({\bm r})
= gN_0 T \sum_{|\omega_n| \le \omega_{\rm cut}} 
     \left\langle \phi_{p\pm}^\ast({\bm k})  
         {\hat{f}}(i{\omega}_n,{\bm r}, {\bm k})  \right\rangle_{\bm k} , 
\label{eq:scD} 
\end{eqnarray} 
where ${\langle} {\dots} {\rangle}_{\bm k}$ indicates Fermi surface average,
$(gN_0)^{-1}=  \ln T +2 T\sum_{0 < \omega_n \le \omega_{\rm cut}}\omega_n^{-1} $, 
and we use $\omega_{\rm cut}=20 k_{\rm B}T_{\rm c}$. 
In Eq. (\ref{eq:scD}), $p$-wave pairing interaction is isotropic in spin space.
For the selfconsistent calculation of the vector potential 
for the internal field $B({\bm r})$,  
we use the current equation
$
\nabla\times \left( \nabla \times {\bm A} \right) 
=
-\frac{2T}{{{\kappa}}^2}  \sum_{0 < \omega_n} 
 \left\langle {\bm v} {\rm Im} \{ g_0  \} 
 \right\rangle_{\bm k}
\quad
$
with the Ginzburg-Landau parameter 
${\kappa}=B_0/\pi k_{\rm B}T_{\rm c}\sqrt{8\pi N_0}$.  
In our calculations, 
we use $\kappa=2.7$ appropriate to $\rm{Sr_2RuO_4}$ as a candidate material for the chiral or helical $p$-wave SC. 
We iterate calculations of Eqs. (\ref{eq:Eil})-(\ref{eq:scD}) for $\omega_n$ 
until we obtain the selfconsistent results of 
${\bm{A}}({\bm{r}})$, ${\hat{\Delta}}({\bm{r}})$ and 
the quasi-classical Green's functions
in the vortex lattice state. 

%%%%%%%%%%%%%%%%%%%%%%%%%%%%%%%%%%%%%%%%%%%%%%%%%%%%%
%-------------------------------------------  Density of State   -------------------------------------------  
%%%%%%%%%%%%%%%%%%%%%%%%%%%%%%%%%%%%%%%%%%%%%%%%%%%%%
In the helical $p$-wave SCs,
$d$-vector is given by ${\bm d}({\bm k}) {\propto}k_x{\hat x}+k_y{\hat y} = {\phi}_{p_+}({\bm k}){\bm d}_- + {\phi}_{p_-}({\bm k}){\bm d}_+$ 
in uniform state at a zero field,
with ${\bm d}_{\pm}({\bm k})={\frac{1}{2}}(1,{\pm}{\mathrm{i}},0)$.
Thus, when we iterate calculations of Eq.(\ref{eq:Eil})-(\ref{eq:scD}),
the initial value of $d$-vector is set to be ${\bm d}({\bm r},{\bm k}) = d({\bm r})(k_x{\hat x}+k_y{\hat y}) $
where $d({\bm r})$ is Abrikosov vortex lattice solution.

Next, using the selfconsistently obtained 
${\bm{A}}({\bm{r}})$ and ${\Delta}({\bm{r}})$, 
we calculate 
${\check{g}}(E\pm{\rm i}\eta,{\bm r},{\bm k})$
for real energy $E$ by
solving Eilenberger eq. (\ref{eq:Eil}) with 
${\rm i}\omega_n \rightarrow E \pm {\rm i}\eta$. 
$\eta$ is a small parameter, and we use $\eta=0.01$ 
in this paper
except for the calculations of distribution in Figs.~\ref{fig04}(d) and \ref{fig04}(e), 
and Figs.~\ref{fig05}(d) and \ref{fig05}(e).
The spin-resolved LDOS $N_{\sigma}(E,{\bm r})$ is given by 
\begin{eqnarray}
N_{\sigma}(E,{\bm r})
=\langle {\rm Re} \{ [ {\hat{g}}(E+{\rm i}\eta,{\bm r},{\bm k}) ]_{{\sigma}{\sigma}} \}\rangle_{\bm k}.
\label{eq:DOS} 
\end{eqnarray}
We define the LDOS $N(E, {\bm r}) =  N_{\downarrow}(E, {\bm r}) + N_{\uparrow}(E, {\bm r}) $,
and spin-polarized LDOS $M(E, {\bm r}) = N_{\downarrow}(E, {\bm r}) - N_{\uparrow}(E, {\bm r})$.

%%%%%%%%%%%%%%%%%%%%%%%%%%%%%%%%%%%%%%%%%%%%%%%%%%%%%
%-------------------------------------------  Calculation Results  -------------------------------------------  
%%%%%%%%%%%%%%%%%%%%%%%%%%%%%%%%%%%%%%%%%%%%%%%%%%%%%
%\section{CALCULATION RESULTS}
%
\section{$H$-dependence of order-parameter}
%%% Fig 1 (H-dependence of oder-parameter ) %%%%%%%%%%%%%% 
\begin{figure}
\begin{center}
\includegraphics[width=8.5cm]{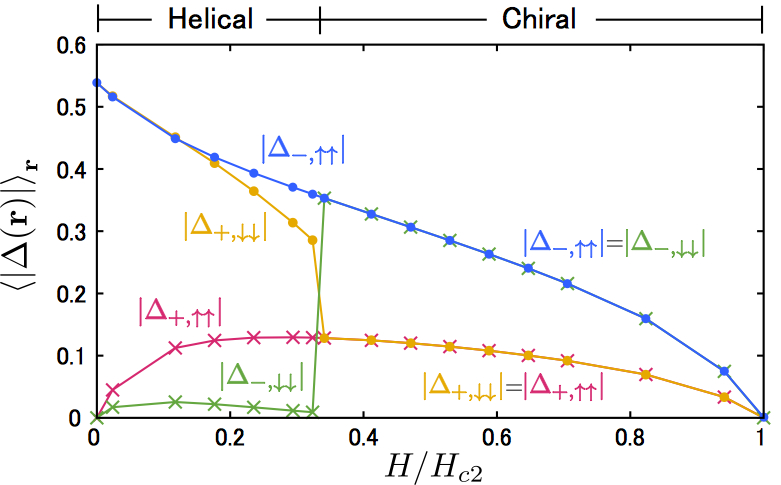}
\end{center}
\vspace{-0.7cm}
\caption{\label{fig01}
$H$-dependence of the spatial average of the order-parameter amplitudes 
${\langle}|{\Delta}_{-, {\downarrow}{\downarrow}}|{\rangle}_{\bm r}$,
${\langle}|{\Delta}_{+, {\downarrow}{\downarrow}}|{\rangle}_{\bm r}$,
${\langle}|{\Delta}_{-, {\uparrow}{\uparrow}}|{\rangle}_{\bm r}$ and
${\langle}|{\Delta}_{+, {\uparrow}{\uparrow}}|{\rangle}_{\bm r}$ 
defined by Eq.~(\ref{eq:op-def}).
The helical $p$-wave state is unstable at $H>0.35H_{c2}$, and 
changes to a chiral $p$-wave state where ${\bm{d}{\perp}{\bm{H}}}$.
}
\end{figure}
%%%%%%%%%%%%%%%%%%%%%%%%%%%%%%%%%

In order to examine the instability of helical $p$-wave state at high $H$,
we show the $H$-dependence of spatial average of the order-parameter amplitude, 
${\langle}|{{\Delta}}_{{\pm},{\sigma}{\sigma}'}({\bm r})|{\rangle}_{\bm r}$ defined by Eq.~(\ref{eq:op-def}) in Fig.~\ref{fig01}.
Using the initial state of helical states,
${\Delta}_{{\downarrow}{\uparrow}}$ and ${\Delta}_{{\uparrow}{\downarrow}}$ components do not appear
in the selfconsistent calculations of our model.
In the vortex state of helical $p$-wave SC at $H<0.35H_{c2}$, 
up-spin pair has a form
${\Delta}_{{\uparrow}{\uparrow}}({\bm r},{\bm k})
={\Delta}_{-,{\uparrow}{\uparrow}}({\bm r}){\phi}_{p_-}({\bm k}) + {\Delta}_{+,{\uparrow}{\uparrow}}({\bm r}){\phi}_{p_+}({\bm k})$
with sub component ${\Delta}_{+,{{\uparrow}{\uparrow}}}({\bm r})$.
The main component ${\Delta}_{-,{{\uparrow}{\uparrow}}}({\bm r})$ has chirality $L_z=-1$, 
anti-parallel to vorticity $W=+1$ as $L_z+W=0$.
The sub component ${\Delta}_{+,{{\uparrow}{\uparrow}}}({\bm r})$ is induced around the vortex core.
Since the local winding number can be a value other than $W=+1$ in the induced components,
the sub component with $L_z=+1$
has inverse winding number $W=-1$ to satisfy the conservation of $L_z+W=0$.~\cite{K.Tanaka3}
According to the previous studies for the vortex state of chiral $p$-wave SC~\cite{Ichioka-chiral, Heeb},
the anti-parallel vortex state ($L_z+W=0$) is stable compared with the parallel vortex state ($L_z+W=+2$)
by the interaction between the chirality and the vorticity.
Therefore,
 the $H$-dependence of ${\langle} | {\Delta}_{-,{\uparrow}{\uparrow}} |{\rangle}_{\bm r}$ and ${\langle} | {\Delta}_{+,{\uparrow}{\uparrow}} |{\rangle}_{\bm r}$
 show same behavior to those for anti-parallel case in a chiral $p$-wave SC~\cite{Ichioka-chiral},
 and the amplitude survives until $H_{c2}$.
 
 On the other hand,
 down-spin pair has a form
 ${\Delta}_{{\downarrow}{\downarrow}}({\bm r},{\bm k})
={\Delta}_{+,{\downarrow}{\downarrow}}({\bm r}){\phi}_{p_+}({\bm k}) + {\Delta}_{-,{\downarrow}{\downarrow}}({\bm r}){\phi}_{p_-}({\bm k})$ at low fields,
 with sub component ${\Delta}_{-,{{\downarrow}{\downarrow}}}({\bm r})$.
 Since the chirality $L_z=+1$ of main ${\Delta}_{+,{\downarrow}{\downarrow}}({\bm r})$ 
 is parallel to vorticity as $L_z+W=+2$,
 ${\Delta}_{{\downarrow}{\downarrow}}({\bm r},{\bm k})$ is rapidly suppressed as a function of $H$, 
 as shown in Fig.~\ref{fig01}.
In addition, at $H{\sim}0.35H_{c2}$,
we find the change of chirality $L_z=+1 {\rightarrow} -1$ in ${\Delta}_{{\downarrow}{\downarrow}}({\bm r},{\bm k})$,
where ${\Delta}_{-,{\downarrow}{\downarrow}}({\bm r},{\bm k})$ changes to be 
main part of  ${\Delta}_{{\downarrow}{\downarrow}}({\bm r},{\bm k})$ 
from the sub component.
At $H>0.35H_{c2}$,
${\langle}|{\Delta}_{-, {\downarrow}{\downarrow}}|{\rangle}_{\bm r}$ is equal to
${\langle}|{\Delta}_{-, {\uparrow}{\uparrow}}|{\rangle}_{\bm r}$
as main components and 
${\langle}|{\Delta}_{+, {\downarrow}{\downarrow}}|{\rangle}_{\bm r}$ is equal to 
${\langle}|{\Delta}_{+, {\uparrow}{\uparrow}}|{\rangle}_{\bm r}$
as sub components,
so that the order-parameter is chiral $p_-$ form.
Even in this chiral state,
${\Delta}_{{\downarrow}{\uparrow}}={\Delta}_{{\uparrow}{\downarrow}}=0$
so that ${\bm d}{\perp}{\bm H}$.
Therefore,
the helical $p$-wave state becomes unstable at high fields
by the effect of vorticity coupling to the chirality,
and changes to a chiral state.

In our model,
we assume that the helical state can appear in the Meissner state $H=0$,
since condensation energy of the helical state is the same as chiral state.
The helical state can be more stable than the chiral state,
if we consider additional mechanism such as weak spin-orbit coupling effect~\cite{Takamatsu_GL}.
Even when very small number of vortices penetrate to the helical $p$-wave SC,
we expect that the helical state can be sustained at the low fields.
With increasing $H$, it becomes metastable state, and finally show instability to the chiral state.
The instability field $H$ can be shifted from our estimation of Fig.~\ref{fig01}.

\section{$H$-dependence of zero-energy spin-polarized DOS and LDOS}
In this section, to find difference of observed quantities between helical and chiral states,
we investigate the characteristic behavior of helical state
under the assumption that the helical $p$-wave state
is sustained at low $H(<0.35H_{c2})$.

\begin{figure}
\begin{center}
\includegraphics[width=8cm]{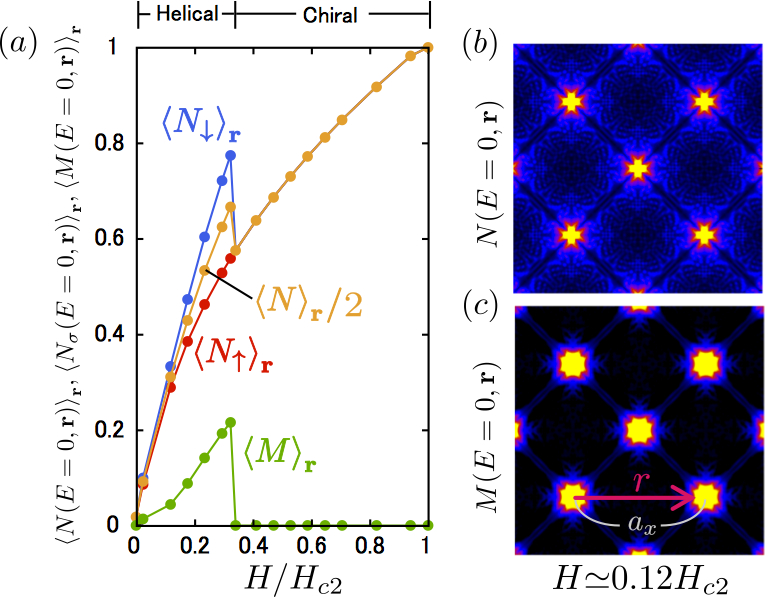}
\end{center}
\vspace{-0.7cm}
\caption{\label{fig02}
(a) $H$-dependence of DOS ${\langle}N(E=0, {\bm r}){\rangle}_{\bm r}/2$,
spin-resolved DOS ${\langle}N_{\sigma}(E=0, {\bm r}){\rangle}_{\bm r}$ and
spin-polarized DOS  ${\langle}M(E=0, {\bm r}){\rangle}_{\bm r}$.
The distributions of zero-energy (b) LDOS $N(E=0, {\bm r}){\le}3$ and
({c}) spin-polarized LDOS $M (E=0, {\bm r}){\le}0.3$
at $H{\simeq}0.12H_{c2}$.
The brighter region indicates the large value of $N$ or $M$.
}
\end{figure}

First,
we study the $H$-dependence of the zero-energy DOS ${\langle}N(E=0, {\bm r}){\rangle}_{\bm r}$,
the zero-energy spin-resolved DOS ${\langle}N_{\sigma}(E=0, {\bm r}){\rangle}_{\bm r}$
and the zero-energy spin-polarized DOS ${\langle}M(E=0, {\bm r}){\rangle}_{\bm r}$.
As shown in Fig.~\ref{fig02}(a), 
the $H$-dependence of ${\langle}N_{{\uparrow}}(E=0, {\bm r}){\rangle}_{\bm r}$ shows the typical behavior, 
which is same behavior in the anti-parallel vortex state of chiral $p$-wave SC~\cite{Ichioka-chiral}.
On the other hand,
the $H$-dependence of ${\langle}N_{\downarrow}(E=0, {\bm r}){\rangle}_{\bm r}$ at $H<0.35H_{c2}$
is larger than 
${\langle}N_{\uparrow}(E=0, {\bm r}){\rangle}_{\bm r}$.
At $H>0.35H_{c2}$,
since 
${\Delta}_{{{\downarrow}{\downarrow}}}$ and ${\Delta}_{{{\uparrow}{\uparrow}}}$ have same chirality,
${\langle}N_{\downarrow}(E=0, {\bm r}){\rangle}_{\bm r}$=${\langle}N_{\uparrow}(E=0, {\bm r}){\rangle}_{\bm r}$.
Here,
contributions of the Zeeman effect are absent since ${\bm d}{\perp}{\bm H}$.
As a result,
the $H$-dependence of DOS ${\langle}N(E=0, {\bm r}){\rangle}_{\bm r}$ 
shows a jump 
when the helical state becomes unstable in Fig.~\ref{fig02}(a).
The jump behavior may be observed by the low temperature specific heat measurement.
When the instability field shifts into high (low) $H$, 
the jump of specific heat becomes larger (smaller).

The $H$-dependence of ${\langle}M(E=0, {\bm r}){\rangle}_{\bm r}$ at low fields
has a finite value and shows increasing behavior, reflecting the ${\langle}N_{\downarrow}(E=0, {\bm r}){\rangle}_{\bm r}$ behavior
in Fig.~\ref{fig02}(a).
And, it jumps to zero when the helical state becomes unstable. 
At high fields as the vortex state of chiral $p$-wave SC,
where ${\Delta}_{{\downarrow}{\downarrow}}={\Delta}_{{\uparrow}{\uparrow}}$, 
$M$ vanishes.
This $H$-dependence of $M$ is the unique behavior of the helical $p$-wave state.
In addition,
Figs.~\ref{fig02}(b) and~\ref{fig02}({c}) show
the LDOS and spin-polarized LDOS distributions at a low field $H{\simeq}0.12H_{c2}$, 
which have large amplitudes around the vortex core.
Since the zero energy state localized around the vortex core is Majorana state in the chiral and helical SCs,
Fig.~\ref{fig02}({c}) shows that the Majorana state is spin-polarized in the helical $p$-wave SCs.
This is another type of spin-polarized zero energy state than that supposed in
 ${\rm{Bi_2Te_3/NbSe_2}}$~\cite{STM_Majonara2} or ${\rm{Cu}}_x{\rm{Bi_2Si_3}}$~\cite{Nagai}.

%%% Fig 2 (H-dependence of DOS ) %%%%%%%%%%%%%% 
%%%%%%%%%%%%%%%%%%%%%%%%%%%%%%%%%

\begin{figure}
\begin{center}
\includegraphics[width=7.5cm]{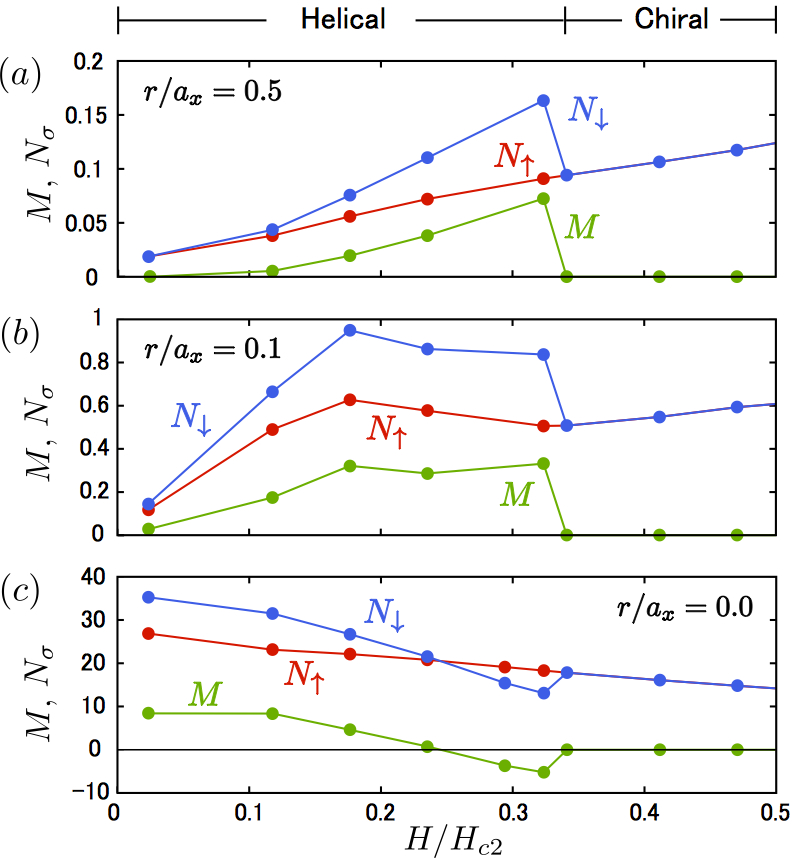}
\end{center}
\vspace{-0.7cm}
\caption{\label{fig03}
(a), (b), ({c})
$H$-dependence of
spin-resolved LDOS $N_{\sigma}(E=0, {\bm r})$ and
spin-polarized LDOS  $M(E=0, {\bm r})$ at radius $r/a_x=$ 0.5, 0.1, 0.0
from the vortex center along the NNN direction, respectively.
$a_x$ is NNN intervortex distance.
}
\end{figure}

Next,
we present the structure of spin-polarized LDOS $M(E, {\bm r})$ at low fields
to study the properties of the vortex state of helical $p$-wave SC.
Figure~\ref{fig03} presents the $H$-dependence of  $M(E=0, {\bm r})$ and $N_{\sigma}(E=0, {\bm r})$ 
at some positions on a line between next-nearest-neighbor (NNN) vortices at $H<0.5H_{c2}$.
At $r/a_x=0.5$ which is midpoint of between NNN vortices, 
 $N_{\downarrow}(E=0, H)>N_{\uparrow}(E=0, H)$ and their magnitudes are small 
and monotonically increase as a function of $H$.
On the other hand, at the vortex core region in Figs.~\ref{fig03}(b) and \ref{fig03}({c}),
$M(E=0, {\bm{r}})$ shows a large amplitude 
at some fields in the helical state.
In particular,
at the vortex center in Fig.~\ref{fig03}({c}),
$M(E=0, {\bm r})$ at $H/H_{c2}{\simeq}$0.02 shows much larger value than the normal state DOS($=1$), 
while it monotonically decrease with raising $H$.
These large values of $M(E=0, {\bm r})$ may be observed by the spin-polarized STM measurement.

%%% Fig 3 (H-dependence of LDOS ) %%%%%%%%%%%%%% 
%%%%%%%%%%%%%%%%%%%%%%%%%%%%%%%%%

\section{$E$-dependence of spin-polarized LDOS}
\begin{figure}
\vspace{0.7cm}
\begin{center}
\includegraphics[width=8.8cm]{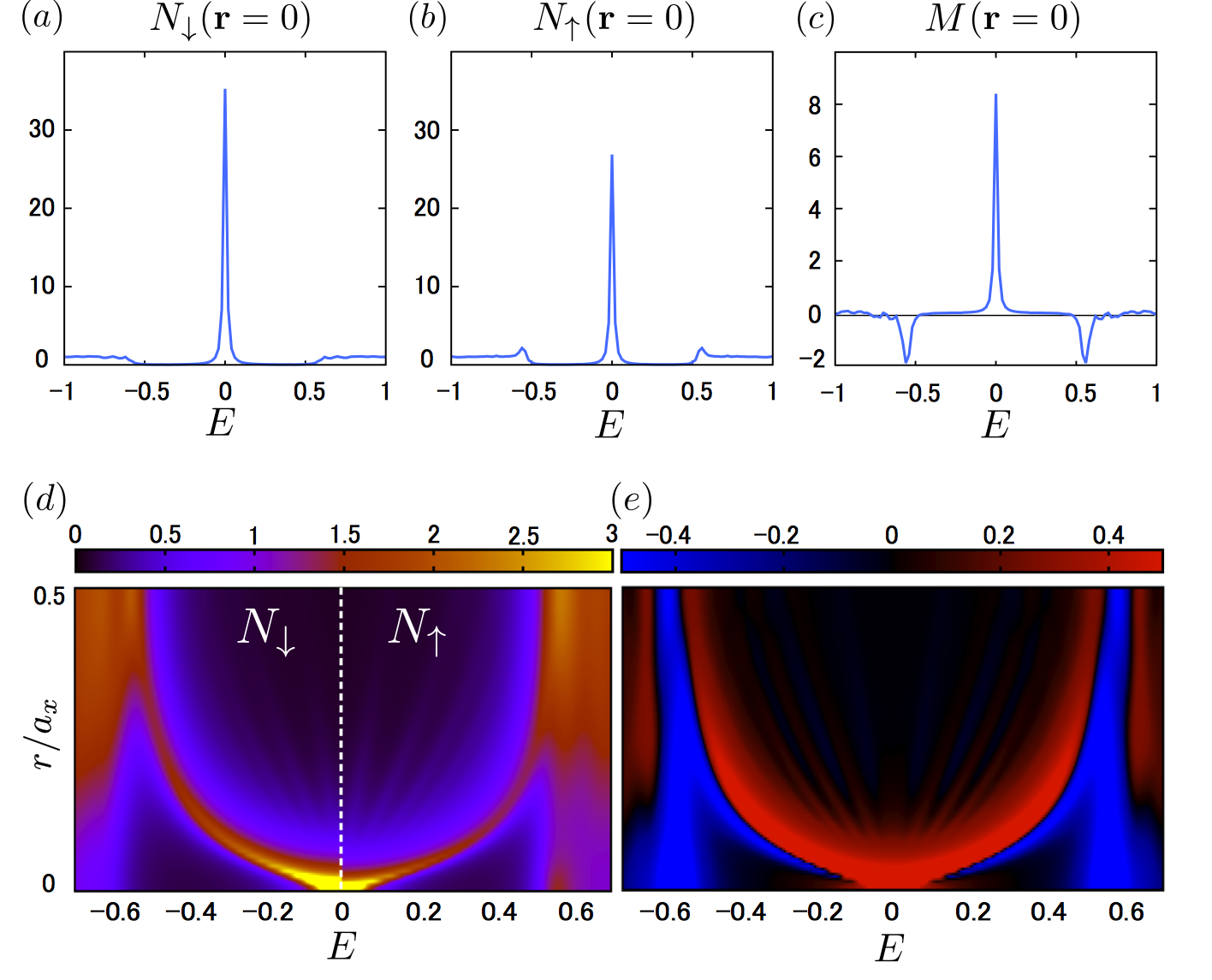}
\end{center}
\vspace{-0.7cm}
\caption{\label{fig04}
(a), (b), ({c}) $E$-dependence of spin-resolved LDOS $N_{\downarrow}$, $N_{\uparrow}$ and 
spin-polarized LDOS $M$
at the vortex center
at $H/H_{c2}{\simeq}$0.02, respectively.
(d), (e) $E$-dependence of $N_{\sigma}(E,{\bm r})$ for ${\sigma}={\downarrow}, {\uparrow}$, and $M(E,{\bm r})$
as a function of radius $r/a_x$ from the vortex center along the NNN direction 
at $H/H_{c2}{\simeq}$0.02, respectively.  
$N_\sigma(-E,r)=N_\sigma(E,r)$. 
In (d) and (e), we use ${\eta}=0.03$.
}
\end{figure}

Finally,
we study the $E$- and $r$-dependences of $N_{\sigma}(E, {\bm r})$ and $M(E, {\bm r})$
in order to investigate the behavior of LDOS spectrum of spin-polarized STM/STS measurement.
When $N_{\uparrow}(E, {\bm r}={\bm 0})$ is compared with $N_{\downarrow}(E, {\bm r}={\bm 0})$ at a low field $H{\simeq}0.02H_{c2}$,
shown in Figs.~\ref{fig04}(a)-({c}),
the height of zero-energy peak in $N_{\uparrow}(E, {\bm r}={\bm 0})$ is smaller,
and instead the gap edges at $E{\sim}{\pm}0.5$ have small peak.
Thus, $M(E, {\bm r}={\bm 0})$ is positive at $E=0$, and negative at $E{\sim}{\pm}0.5$.
These weights cancel each other, 
so that total spin polarization ${\int_{-{\infty}}^0 M(E,{\bm r})}dE = 0$. 
This condition can be extended to finite $T$ as ${\int_{-{\infty}}^{\infty} M(E,{\bm r})}F(E,T)dE = 0$
with Fermi distribution function $F(E,T)$ since $M(E, {\bm r})$ is even function of $E$.
The absence of total spin polarization corresponds to the fact
that Knight shift is invariant in the helical $p$-wave state, where ${\bm d}{\perp}{\bm H}$.
To observe the spin-polarized LDOS in the helical state,
we have to perform $E$-resolved observation such as spin-polarized STM/STS.
The $r$-dependence of spectra $N_\sigma(E,r)$ and $M(E,r)$ are presented in Figs.~\ref{fig04}(d) and \ref{fig04}(e), respectively. 
When we focus on the dispersion curve of brighter region in Fig.~\ref{fig04}(d),
the zero-energy peak at $r=0$ evolves toward  the gap-edge with increasing $r$. 
Since the zero-energy vortex bound state connects with the gap-edge state at smaller $r$ for $N_{\uparrow}$ than $N_{\downarrow}$,
the effective vortex core radius is smaller for $N_{\uparrow}$. 
Therefore,
in $N_{\uparrow}$, the peaks of the gap edge ($E {\sim} {\pm} 0.5$) outside vortices
can extend until the vortex center, as shown in Fig.~\ref{fig04}(b).  
In Fig.~\ref{fig04}(e), we see that the spin-polarized state appears near 
the dispersion curve of vortex bound state extending from the Majorana zero mode, 
in addition to gap edges. 

%%% Fig 4 (E-dependence of LDOS ) %%%%%%%%%%%%%% 
%%%%%%%%%%%%%%%%%%%%%%%%%%%%%%%%%

\begin{figure}
\vspace{0.7cm}
\begin{center}
\includegraphics[width=8.6cm]{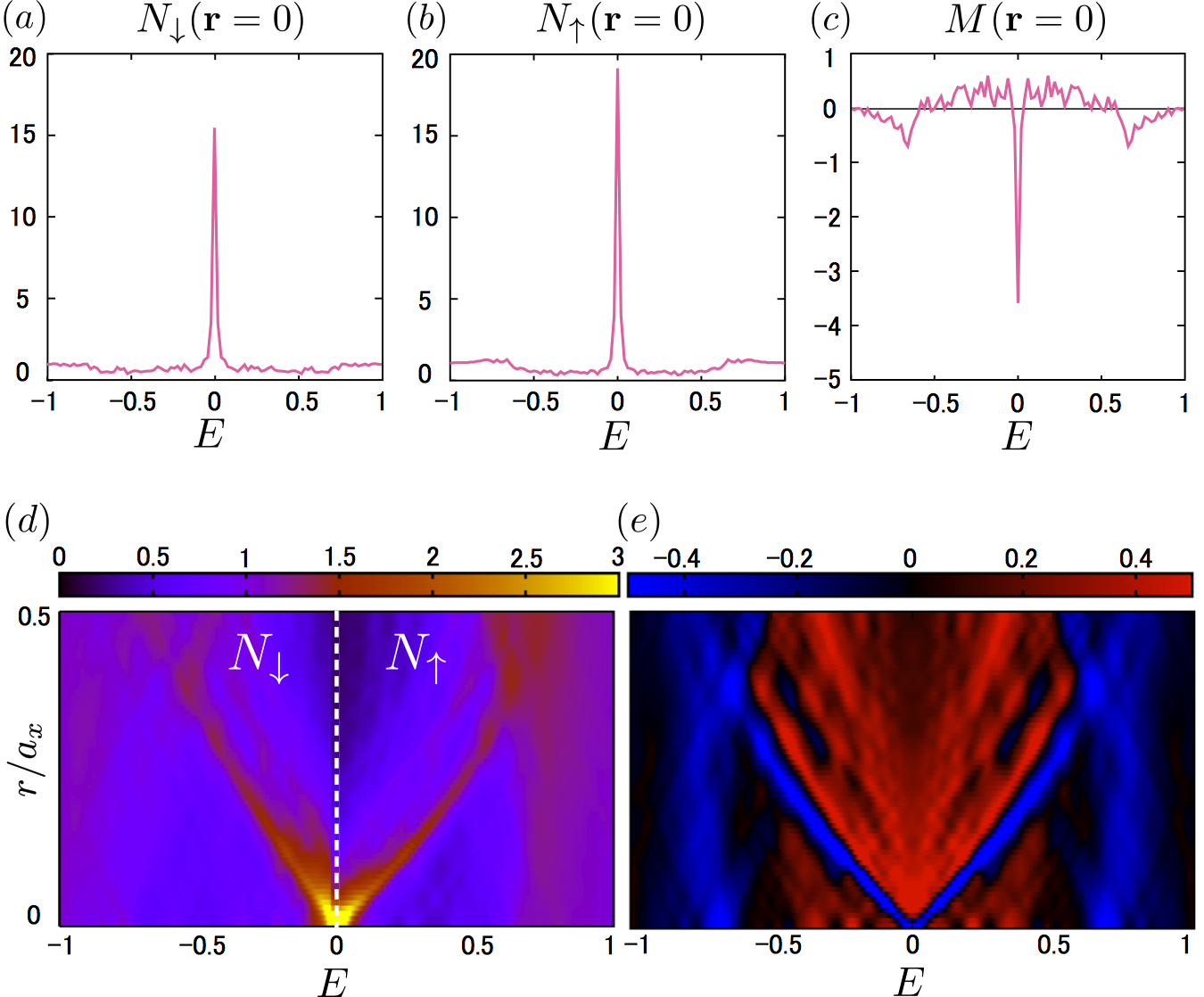}
\end{center}
\vspace{-0.7cm}
\caption{\label{fig05}
(a), (b), ({c}) $E$-dependence of spin-resolved LDOS $N_{\downarrow}$, $N_{\uparrow}$ and 
spin-polarized LDOS $M$
at the vortex center
at $H/H_{c2}{\simeq}$0.29, respectively.
(d), (e) $E$-dependence of $N_{\sigma}(E,{\bm r})$ for ${\sigma}={\downarrow}, {\uparrow}$, and $M(E,{\bm r})$
as a function of radius $r/a_x$ from the vortex center along the NNN direction 
at $H/H_{c2}{\simeq}$0.29, respectively.  
$N_\sigma(-E,r)=N_\sigma(E,r)$. 
In (d) and (e), we use ${\eta}=0.03$.
}
\end{figure}

Moreover,
we show the $E$- and $r$-dependences of $N_{\sigma}(E, {\bm r})$ and $M(E, {\bm r})$ at a higher field $H{\simeq}0.29H_{c2}$,
considering that the helical $p$-wave state is still sustained at higher $H$.
In Figs.~\ref{fig05}(a)-({c}),
the hight of zero-energy peak of $N_{\uparrow}$ is larger than  $N_{\downarrow}$,
resulted in negative $M(E=0,{\bm r}={\bm 0})$.
To compensate negative value at $E=0$ and at the gap edge,  
$M(E,r=0)$ becomes positive for in-gap states for $0<|E|<0.5$. 
As shown in Figs.~\ref{fig05}(d) and \ref{fig05}(e),
since the down-spin's in-gap states have a larger value compared with the up-spin states,
$M(E,{\bm r})$ has finite distributions at $0<|E|<0.5$ 
even far from dispersion curve of bound state.

%%% Fig 5 (E-dependence of LDOS ) %%%%%%%%%%%%%% 
%%%%%%%%%%%%%%%%%%%%%%%%%%%%%%%%%

%%%%%%%%%%%%%%%%%%%%%%%%%%%%%%%%%%%%%%%%%%%%%%%%%%%%%
%-------------------------------------------  Summary  -------------------------------------------  
%%%%%%%%%%%%%%%%%%%%%%%%%%%%%%%%%%%%%%%%%%%%%%%%%%%%%
\section{Summary}
We studied the vortex state of helical $p$-wave SCs based on the quasi-classical Eilenberger theory.
We confirmed the instability of the helical $p$-wave state at high fields
and that the spin-polarized LDOS $M(E, {\bm r})$ appears even when Knight shift does not change.
This is because the vorticity couples to the chirality of up- or down-spin pair of helical state.
In addition,
we found that the magnetic field dependence of zero-energy DOS shows a jump 
when the helical state becomes unstable.
This jump behavior may be observed by the low temperature specific heat measurement.
In order to identify the helical $p$-wave state at low fields,
we investigated the structure of the zero-energy $M(E=0, {\bm r})$ in the vortex states.
In particular, at the vortex center,
the value of $M(E=0, {\bm r}=0)$ at a low field $H/H_{c2}{\simeq}$0.02 
shows much larger value than the normal state DOS, 
while it monotonically decrease with raising field.
Moreover,
we present the $E$- and $r$-dependences of the spin-resolved LDOS
$N_{\downarrow}(E,{\bm r})$, $N_{\uparrow}(E,{\bm r})$ and $M(E,{\bm r})$ 
in the vortex state.
We hope that these theoretical calculation results of spin-polarized LDOS will be examined,
and will be used for detecting the spin-polarized Majorana zero-energy modes 
by the spin-polarized STM/STS measurement.

\begin{acknowledgments}
This work was supported by JSPS KAKENHI Grant Number JP16J05824.
\end{acknowledgments}

%--------------------------

\end{document}